# Second Harmonic Generation and Nonlinear Frequency Conversion in Photonic Time-Crystals

Noa Konforty[1,2], Moshe-Ishay Cohen[1,2], Ohad Segal[2,3], Yonatan Plotnik[2], and Mordechai Segev[1,2,3]

[1] Physics Department, Technion – Israel Institute of Technology, Haifa 32000, Israel

[2] Solid State Institute, Technion – Israel Institute of Technology, Haifa 32000, Israel

[3] Department of Electrical and Computer Engineering, Technion – Israel Institute of Technology, Haifa 32000, Israel

## Abstract

We study the nonlinear process of second harmonic generation in photonic time-crystals, materials with refractive index that varies abruptly and periodically in time, and obtain the phase matching condition for this process. We find conditions for which the second harmonic generation is highly enhanced even in the absence of phase matching, governed by the exponential growth of the modes residing in the momentum gap of the photonic time crystal. Additionally, under these conditions, a cascade of higher order harmonics is generated at growing exponential rates. The process is robust, with no requirement for phase-matching, the presence of a resonance or a threshold, drawing energy from the modulation.

The exploration of epsilon-near-zero materials presents new opportunities to create time-interfaces with large and abrupt changes in the refraction index [1–14]. The recent advancements in theory and experiments have drawn increasing attention to Photonic Time-Crystals (PTCs) [15–19]. PTCs are materials with refractive index that undergoes substantial periodic variations on the time-scales of a single optical cycle. They exhibit dispersion relation displaying momentum bands, separated by momentum gaps wherein the electromagnetic (EM) modes are exponentially growing (or decaying) in time [15,20–23]. The study of PTCs is introducing new avenues for shaping light-matter interactions, relevant both for lasing technologies as well as for quantum technologies, offering new sources of entangled states [16,24–27]. The special dispersion relation in PTCs arises from interference between multiple time-reflected and time-refracted EM waves which are generated from the abrupt variations to the refractive index [22,23,28–39]. The exponentially growing modes associated with the momentum gap are possible because the time-symmetry is broken by the modulation of the refractive index. Namely, the growing modes extract energy from the index modulation, while the decaying modes transfer energy to it. Importantly, this energy exchange between the gap modes and the index modulation is non-resonant, hence it can support numerous visionary ideas such as lasers that do not rely on any atomic resonance [27], non-resonant creation of pairs of entangled photons [25,27], etc. Recently, studies of the nonlinear phenomenon of solitons in nonlinear PTCs has demonstrated the unique properties of the momentum gaps in PTCs, and predicted the existence of superluminal $k$-gap solitons [40].

The unusual dispersion relation in photonic time-crystals was thus far not utilized in the context of nonlinear frequency conversion - a core concept in nonlinear optics and in fact the first nonlinear optical phenomenon to be discovered [41]. Nonlinear frequency conversion is strongly affected by phase-matching [42,43]. Generally, phase matching depends on the dispersion in the medium and often does not occur naturally. Over the years, many methods for phase-matching have been explored, the most important ones being birefringence phase-matching [42] and quasi-phase-matching [44,45]. The most basic and commonly used nonlinear frequency conversion process is second harmonic generation (SHG), where due to the non-centrosymmetric structure of certain materials, a propagating wave with frequency $\omega$ excites the spatially-asymmetric dipoles in the medium, which, in turn, emit radiation at frequency $2\omega$ [41,44,46,47].

Here, we explore second harmonic generation in photonic time-crystals, and find that SHG can be exponentially enhanced when momentum gap modes are involved. We find the phase matching conditions for the Floquet modes associated with the momentum bands and the momentum gaps of the PTC. We show how the momentum bandgaps in PTCs can enable exponentially amplified SHG even without phase matching, extracting energy from the modulation of the refractive index. The amplification in the gaps is non-resonant, and without any threshold requirements. Moreover, we observe a dramatic cascading effect of the emergence of higher order harmonics with wavenumbers $nk_0$, where $k_0$ is the wavenumber of the fundamental mode (that resides within the momentum gap of the PTC), and n is an integer. Each harmonic is growing exponentially, with no saturation effect, drawing the energy from the modulation of the PTC. Finally, we discuss how the exponential amplification of the SHG process can pave the way towards designing momentum bandgaps at multiple wavelengths simultaneously by employing cascaded $\chi^{(2)}$ processes, and envision new physical mechanisms for high-harmonic generation in solids and exploiting them for ultrashort laser pulses.

For simplicity, we consider a nonlinear coefficient $\chi^{(2)}$ that is constant in time and does not depend on frequency, as is always the case far from atomic resonances or bandgaps in solids. For clarity, we define the frequency and momentum of the original wave as $\omega_0$ and $k_0$, respectively, and the frequency and momentum of the generated wave as $\omega_{SH}$ and $k_{SH}$. It is important to note that the true frequency of each of the waves is well-defined only when the medium is stationary in time, because the frequency varies during the modulation.

In conventional SHG process, we consider a finite nonlinear medium, breaking homogeneity in space (at the entrance and exit planes) but time-translation symmetry is conserved, hence the conserved quantity is frequency. Thus, the solution to the SHG process is a time-harmonic wave with a well-defined single frequency $\omega_{SH} = 2\omega_0$. The generated wave has a spatially-varying envelope, that changes as the SH mode draws energy from the fundamental mode (pump). The phase-matching condition ensures efficient transfer of energy from the fundamental mode to the SH mode, in which case its spatial envelope grows in the medium. If the phase matching condition is not met, energy is transferred back and forth between the fundamental and SH modes, and the spatial envelope oscillates in space.

The logic in the SHG process in time-varying media is different. In a time-varying medium, it is natural to work with spatially-homogeneous boundary conditions, which implies conservation of momentum. Hence, the solution for the nonlinear process is a spatially-harmonic wave with a well-defined wavenumber $k_{SH} = 2k_0$. The generated SH wave now has a time-varying envelope (in analogy to conventional SHG in stationary media). Here, phase-matching conditions will depend on the temporal frequencies of the fundamental and SH modes, as we show below for PTCs.

For this reason, throughout this work we consider spatially-homogeneous boundary conditions. The medium is stationary until the moment $t = 0$, then a PTC is established by modulating the refractive index periodically in time, $n(t) = n_0 + n_1(t)$, with time periodicity $T$, such that $n_1(t) = n_1(t + T)$. The modulation stops after $N$ cycles, at $t = NT$, such that $N$ is large (at least $N=20$), as sketched in Fig. 1A. Due to our spatially-homogeneous boundary conditions, we will refer to the fundamental and SH modes by their well-defined wavenumbers: $k, 2k$.

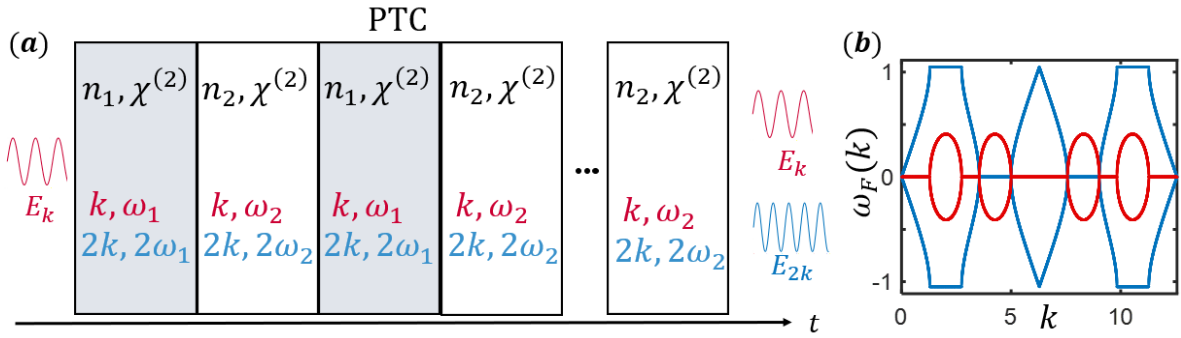

Figure 1: (a) Sketch of the system. A wave with wavenumber $k$ enters a nonlinear spatially-homogeneous PTC. The nonlinearity generates a second harmonic wave, with wavenumber $2k$. (b) Band structure of a linear PTC. In blue is the real component of the Floquet frequency $\omega_F(k)$, and in red is the imaginary component.

Following the modulation defining the PTC, the EM waves in a linear medium have the form of Floquet solutions:

$$E_k(z,t) = u(t)e^{i(\omega_F(k)t - kz)} \; ; \; E_{2k}(z,t) = v(t)e^{i(\omega_F(2k)t - 2kz)} \quad (1)$$

The solutions are plane-waves, multiplied by periodic functions with the periodicity of the modulation, $T$, i.e., $v(t) = v(t + T)$, $u(t) = u(t + T)$. The frequencies $\omega_F(k)$, $\omega_F(2k)$ are the Floquet frequencies, determined from the band structure of the PTC (Fig. 1B), similar to the momentum vector of Bloch modes in spatial crystals. The Floquet frequencies differ from the true frequencies of a time-harmonic mode in a stationary medium, that are derived from the static refractive index of the material. In the absence of nonlinearity, $u(t)$ and $v(t)$ are decoupled from each other.

For the nonlinear case of SHG, the electric field must satisfy the nonlinear wave equation

$$\nabla^2 E = \mu \left( \frac{\partial^2 \varepsilon(t)}{\partial t^2} E + \varepsilon(t) \frac{\partial^2 E}{\partial t^2} + 2 \frac{\partial \varepsilon(t)}{\partial t} \frac{\partial E}{\partial t} \right) + \mu \frac{\partial^2}{\partial t^2}(\chi^{(2)} E^2) \quad (2)$$

Under the non-depleted pump approximation for the fundamental mode with wavenumber $k$, the amplitude of the fundamental mode, $A$, does not depend on time, hence the solution is of the form:

$$E_k(z,t) = Au(t)e^{i(\omega_F(k)t - kz)} + c.c. \quad ; \quad E_{2k}(z,t) = B(t)v(t)e^{i(\omega_F(2k)t - 2kz)} + c.c. \quad (3)$$

Next, we substitute the solution form into Eq. 2 and employ the slowly time-varying envelope approximation $|\ddot{B}| \ll |\omega_F \dot{B}|, |\dot{v}\dot{B}|$. We keep only spatially-synchronized terms containing the wavenumber $2k$. We obtain the following result, denoting a new function $f(t)$, and the constant $\Delta\omega$, representing the phase mismatch:

$$\dot{B} = -A^2 \chi^{(2)} f(t) e^{-i\Delta\omega t} \quad ; \quad f(t) \triangleq \frac{u\ddot{u} + \dot{u}^2 - 2\omega_F^2(k)u^2 + 4i\omega_F(k)u\dot{u}}{\varepsilon(\dot{v} + i\omega_F(2k)v) + \dot{\varepsilon}v} \quad ; \quad (4)$$

$$\Delta\omega \triangleq \omega_F(2k) - 2\omega_F(k)$$

Since $f(t)$ is composed of functions with time periodicity $T$, we expand it as a Fourier series. Defining $\Omega \triangleq \frac{2\pi}{T}$, $f(t) = \sum_{n\in\mathbb{Z}} a_n e^{i\Omega nt}$, we obtain the amplitude of the second harmonic, as a function of time $t$:

$$B(t) = -2iA^2 \chi^{(2)} \sum_n a_n \left( \frac{e^{i(\Omega n - \Delta\omega)t} - 1}{i(\Omega n - \Delta\omega)} \right) \quad (5)$$

The solution for the SH field is:

$$E_{2k}(t) = -2iA^2\chi^{(2)} \sum_n a_n \left(\frac{e^{i(\Omega n - \Delta\omega)t} - 1}{i(\Omega n - \Delta\omega)}\right) v(t) e^{i(\omega_F(2k)t - 2kz)} + c.c. \quad (6)$$

$|B(t)|^2$ is related to the amount of energy transferred from the fundamental field (pump) to the SH field. In the conventional SHG derivation in stationary nonlinear media, efficient SHG requires phase-matching $\Delta k = 2k(\omega) - k(2\omega) = 0$. In that case, the relative phase between the fundamental and the harmonic field is constant $\frac{\pi}{2}$, which is the optimal phase for transfer of energy, and therefore energy is constantly transferred from the fundamental to the harmonic. In our case, if $Re(\Delta\omega) \equiv Re(\omega_F(2k) - 2\omega_F(k)) = n\Omega$ for some integer $n$, we obtain the analogous phase matching condition in the PTC, where the oscillations in $\varepsilon(t)$ enable this Floquet phase matching condition.

We note that phase matching in a PTC depends on the Floquet frequency, and not on the true frequency $\omega(k)$ derived from the refractive index in a stationary medium. The Floquet frequency can be controlled by changing the modulation, and through that the band structure of the PTC can be engineered to phase-match a selected second harmonic.

More exotic cases arise when one or both momenta $k, 2k$ are in the PTC's momentum gap, and the Floquet frequencies have a complex part. The linear PTC modes in the momentum gap have two branches – one with exponentially growing modes ($Im(\omega_F) > 0$), and another with exponentially decaying modes ($Im(\omega_F) < 0$). When we examine the form of $B(t)$, we see that the envelope grows exponentially in time if $Im(\Delta\omega) < 0$, and decreases if $Im(\Delta\omega) > 0$. In those cases, even if $Re(\Delta\omega) \neq 0$ and we have phase mismatch, the dominance of the exponential growth overpowers any modulation from the phase mismatch, and the SH process becomes efficient. Henceforth we consider four generic cases.

Case I: Both the fundamental and the SH modes are in the band and their Floquet frequencies are real. In this case the outcome is analogous to the conventional case of SHG in a stationary medium. If phase matching is obtained, the envelope of the SH field is linearly growing in time, and the intensity is growing parabolically (Fig. 2A). If they are not phase matched, the intensity of the of the SH oscillates periodically with periodicity $2\pi/\Delta\omega$, and the power periodically transferred to and from the pump to the SH (Fig. 2B).

Case II: The wavenumber of the fundamental mode, $k$, is in the band while the SH wavenumber $2k$ is in the momentum gap. Since the PTC's band structure allows for modes with either negative or positive complex Floquet frequencies, this case has modes with $Im(\Delta\omega) < 0$ or $Im(\Delta\omega) > 0$. Consider first the mode with $Im(\Delta\omega) < 0$ which implies $Im(\omega_F(2k)) < 0$, giving rise to exponentially growing $B(t)$. In this case, despite the growth in $B(t)$ representing energy transfer from the pump to the SH, the process is also coupled to the exponentially decreasing mode in the momentum gap, and overall, this mode does not experience exponential growth and does not become a dominant mode. In the other case, where the process begins with the mode with $Im(\omega_F(2k)) > 0$, with decaying power transfer from the fundamental to $B(t)$, the process is coupled to the growing gap mode of the PTC. In reality, one cannot generate a case when only the decaying mode exists, because the interaction with the fundamental will always mixes the two states. Therefore, the SH mode always has the seed it needs to exponentially grow, and the long-term dynamics is dominated by the exponentially growing mode of the SH residing in the momentum gap of the PTC (Fig. 2C).

Case III: This case is the opposite of Case II. Here, the wavenumber $k$ of the fundamental mode is in the gap while the SH's wavenumber $2k$ is in the band. When $Im(\omega_F(k)) > 0$, $B(t)$, which implies that the power transfer from the exponentially growing

$k$ mode to the SH field grows exponentially, with an exponent twice the exponent of the fundamental mode (Fig. 2D).

Case IV: This last case occurs when both the pump and the SH are in the gap. In this case, the dominant exponent is either $e^{Im(2\omega_F(k))}$ or $e^{Im(\omega_F(2k))}$, depending on which one is larger, as shown in equation (6) (Fig. 2E-F).

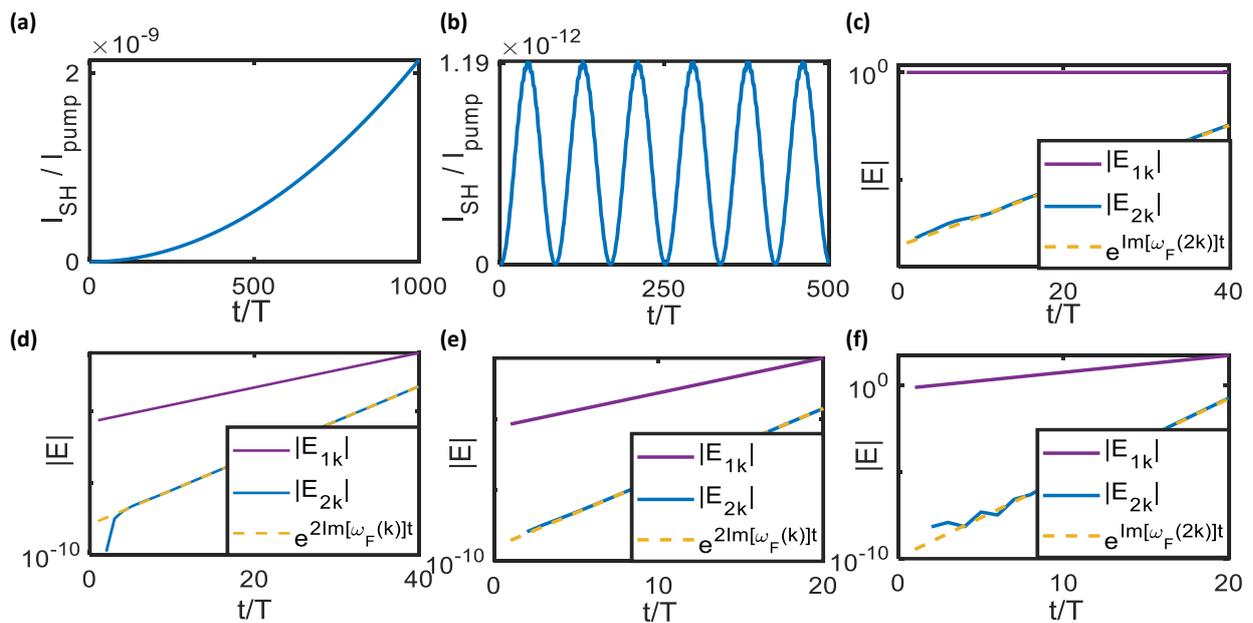

Figure 2: (a,b) **Case I**: Intensity of the SH as a function of time, with both $k$ and $2k$ in the band, with (a) and without (b) phase-matching, respectively. When process is phase matched (a) the SH intensity grows parabolically with time, whereas for a phase-mismatched process (b) the SH intensity oscillates in time. (c) **Case II**: Phase-mismatched SHG with the SH in the momentum gap: The SH grows exponentially despite the phase mismatch. The vertical axis is logarithmic. (d) **Case III:** Phase mismatched SHG with the fundamental in the gap: both the fundamental and second harmonics grow exponentially, despite the phase mismatch. The second harmonic grows with an exponential rate twice as large than the first harmonic, even though its mode belongs to the band. (e,f) **Case IV:** Phase mismatched SHG with both fundamental and SH in gap, where the dominant exponent is the exponent of the fundamental mode (e) and where the dominant exponent is the exponent of the SH mode (f).

Importantly, just like the exponential amplification of gapped modes in PTCs, the exponential amplification of the SH signal is not affected by phase mismatch because the exponential amplification associated with gap modes overcomes the phase-mismatch. This is seemingly similar to OPAs, which can also support amplification without phase-matching.

However, this similarity is misleading, because in OPAs amplification without phase-matching occurs only above a threshold (when the magnitude of the gain coefficient is larger than the phase-mismatch), whereas here the amplification always occurs, without any threshold, even for large phase-mismatch. Moreover, the process here is non-resonant, and takes place for every fundamental signal that is in the bandgap.

Next, we ask what happens when the second harmonic signal becomes strong, due to the exponential amplification. We focus on the case where the wavenumber of the fundamental ($k$) is in the gap and the SH ($2k$) is in the band, hence the SH is growing at a faster rate than the fundamental (Case II). In the conventional SHG process, the fundamental is depleted when a non-negligible fraction of its power is transferred to the SH. However, unlike conventional SHG which is a parametric process and its energy is conserved and the pump is always depleted when the SH becomes stronger, here, the PTC modulation keeps driving energy into the system. Consequently, as the SH gets stronger, we observe a dominant cascading effect: higher order harmonics emerge, and they also grow at exponential rates. For example, a signal with wavenumber $3k$ is generated by a sum-frequency process of the fundamental and SH. Likewise, a signal with wavenumber $4k$ is generated by sum frequency of the SH signal with itself, and so on, until we observe high-order harmonics with wavenumber $nk_o$, where $n$ is an integer. As an example, consider the generation of the 4$^{\text{th}}$ harmonic with wavenumber $4k$, generated by cascading SHG, where the SH ($2k$) arising from SHG of the fundamental at wavenumber $k$ serves as the fundamental for the new $\chi^{(2)}$ process. Consequentially, the fundamental signal for the new process is displaying the same exponential growth as a gapped mode. Therefore, we expect that the new nonlinear signal (with wavenumber $4k$) will grow with an exponential rate that is twice the exponent of the generating signal, i.e., with exponent $e^{Im(4\omega_F(k_0))}$. Overall, each signal with wavenumber $nk_0$ is expected to grow at an exponential rate $e^{Im(n\omega_F(k_0))}$, even though only the fundamental signal $k$ is in the linear PTC's momentum bandgap, as shown in Fig. 3. We emphasize that the power of all the different harmonics may become comparable to the power in the fundamental, yet still - the power carried by the harmonics keeps growing exponentially without any saturation, drawing energy from the modulation of the refractive index.

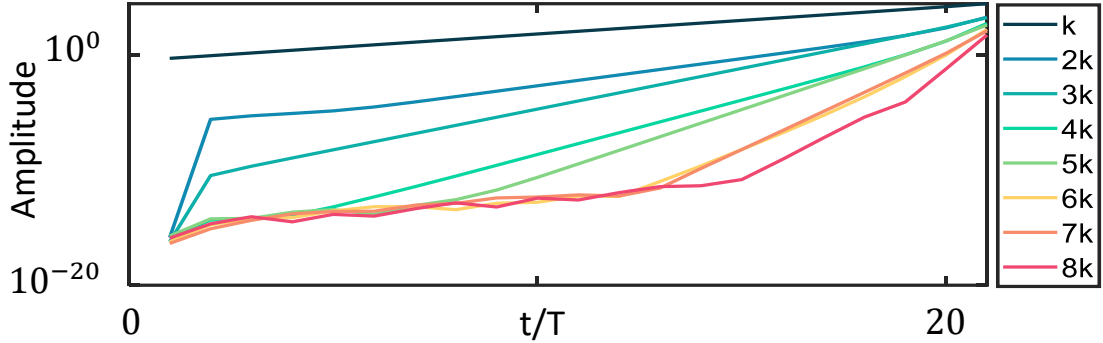

Figure 3: Cascaded harmonics generation: Intensity of modes with different wave numbers as a function of time. The cascading of the harmonics grows exponentially, at increasing exponential rates, with no pump depletion, drawing their energy from the modulation creating the PTC.

The nonlinear cascade of harmonics exemplifies how the band structure of the PTC changes dramatically when we introduce nonlinear optics. Considering for example a PTC with one bandgap around $k$, and a nonlinear coefficient $\chi^{(2)} \neq 0$, in the presence of an EM mode $k$, we expect now all modes with wavenumber $nk_0$ to behave as gapped modes that grow exponentially. This could have many applications. For example, it could pave the way for introducing momentum bandgaps at optical frequencies, by harvesting nonlinear $\chi^{(2)}$ or $\chi^{(3)}$ processes, which have not yet been accesses experimentally, by exploiting nonlinear $\chi^{(2)}$ or $\chi^{(3)}$ processes.

To conclude, we studied the generation of a second harmonic wave in a photonic time-crystal, and found that the phase matching condition depends on the Floquet frequencies in the band structure of the photonic time crystal. Recalling that the band-structure of a PTC is shaped by the time-modulation, this exemplifies the potential of dispersion engineering in time-varying materials, specifically in shaping the nonlinear processes. Moreover, when one of the interacting waves is associated with a momentum bandgap, we observe exponential amplification of the second harmonic, followed by a cascade of higher harmonics growing at faster exponential rate. The process does not require phase-matching or a resonance of any kind, and lacks any threshold. In fact, in the process of nonlinear frequency conversion in PTCs the nonlinearity acts as a mediator, facilitating energy flow from the modulation of the refractive index to the various harmonics, effectively changing the PTC's band structure. The cascading effect can pave the way for new experimental methods to achieve a momentum bandgap in the optical regime of EM waves. Moreover, the creation of an exponentially growing frequency comb in this process implies that a nonlinear PTC can be used to generate ultrashort laser pulses. Physically, the order of the harmonics generated in this cascading process can be high, limited only by the material response (via the value of $\chi^{(2)}$) and by the rate of the modulation of the refractive index. As such, it is possible, in principle, that this could serve as a new process for High Harmonic Generation in solids.

The concepts presented here examined SHG, but obviously they apply to all $\chi^{(2)}$ processes, and higher order processes ($\chi^{(3)}$ and higher) can be formulated in the same way. In the broad context of nonlinear frequency conversion in time-varying media, several general intriguing questions arise. For example, thus far we considered a fixed value for the nonlinear coefficient $\chi^{(2)}$, but when the modulation is very strong, the atomic potential is likely to be affected, and therefore the nonlinearities will also present time-periodic dependence. Moreover, following our recent experiments [7], we know we can induce refractive index changes of order unity within 6fsec. Can we also induce large changes to nonlinear coefficients of solid materials on the single-optical-cycle timescale? Along these lines, new works have already started exploring epsilon-near-zero materials as interesting candidates for experimental nonlinear time-varying optics [48,49,50].

This work was supported by the Israel Science Foundation through the MAPATS program, and by the US Air Force Office for Scientific Research, AFOSR.

**References:**

[1] N. Kinsey, C. DeVault, J. Kim, M. Ferrera, V. M. Shalaev, and A. Boltasseva, Epsilon-near-zero Al-doped ZnO for ultrafast switching at telecom wavelengths, Optica **2**, 616 (2015).

[2] L. Caspani et al., Enhanced Nonlinear Refractive Index in ϵ-Near-Zero Materials, Phys Rev Lett **116**, (2016).

[3] M. Z. Alam, I. De Leon, and R. W. Boyd, Large optical nonlinearity of indium tin oxide in its epsilon-near-zero region, Science (1979) **352**, 795 (2016).

[4] O. Reshef, I. De Leon, M. Z. Alam, and R. W. Boyd, *Nonlinear Optical Effects in Epsilon-near-Zero Media*, Nature Reviews Materials.

[5] V. Bruno, S. Vezzoli, C. De Vault, E. Carnemolla, M. Ferrera, A. Boltasseva, V. M. Shalaev, D. Faccio, and M. Clerici, Broad frequency shift of parametric processes in epsilon-near-zero time-varying media, Applied Sciences (Switzerland) **10**, (2020).

[6] Y. Zhou, M. Z. Alam, M. Karimi, J. Upham, O. Reshef, C. Liu, A. E. Willner, and R. W. Boyd, Broadband frequency translation through time refraction in an epsilon-near-zero material, Nat Commun **11**, (2020).

[7] E. Lustig et al., Time-refraction optics with single cycle modulation, Nanophotonics **12**, 2221 (2023).

[8] S. Saha, O. Segal, C. Fruhling, E. Lustig, M. Segev, A. Boltasseva, and V. M. Shalaev, Photonic time crystals: a materials perspective Invited, Opt. Express **31**, 8267 (2023).

[9] R. Tirole, S. Vezzoli, E. Galiffi, I. Robertson, D. Maurice, B. Tilmann, S. A. Maier, J. B. Pendry, and R. Sapienza, Double-Slit Time Diffraction at Optical Frequencies, n.d.

[10] R. Maas, J. Parsons, N. Engheta, and A. Polman, Experimental realization of an epsilon-near-zero metamaterial at visible wavelengths, Nat Photonics **7**, 907 (2013).

[11] L. Caspani et al., Enhanced Nonlinear Refractive Index in ϵ-Near-Zero Materials, Phys Rev Lett **116**, 233901 (2016).

[12] V. Bruno et al., Negative Refraction in Time-Varying Strongly Coupled Plasmonic-Antenna–Epsilon-Near-Zero Systems, Phys Rev Lett **124**, 043902 (2020).

[13] R. Tirole et al., Saturable Time-Varying Mirror Based on an Epsilon-Near-Zero Material, Phys Rev Appl **18**, 054067 (2022).

[14] X. Wang, M. S. Mirmoosa, V. S. Asadchy, C. Rockstuhl, S. Fan, and S. A. Tretyakov, Metasurface-based realization of photonic time crystals, Sci Adv **9**, eadg7541 (2023).

[15] E. Lustig, Y. Sharabi, and M. Segev, Topological aspects of photonic time crystals, Optica **5**, 1390 (2018).

[16] E. Lustig, O. Segal, S. Saha, C. Fruhling, V. M. Shalaev, A. Boltasseva, and M. Segev, Photonic time-crystals - fundamental concepts Invited, Opt. Express **31**, 9165 (2023).

[17] X. Wang, M. S. Mirmoosa, V. S. Asadchy, C. Rockstuhl, S. Fan, and S. A. Tretyakov, Metasurface-Based Realization of Photonic Time Crystals, (2022).

[18]  J. S. Martínez-Romero and P. Halevi, Parametric resonances in a temporal photonic crystal slab, Phys Rev A (Coll Park) **98**, 053852 (2018).

[19]  N. Wang, Z.-Q. Zhang, and C. T. Chan, Photonic Floquet media with a complex time-periodic permittivity, Phys Rev B **98**, 085142 (2018).

[20]  J. R. Reyes-Ayona and P. Halevi, Observation of genuine wave vector (k or β) gap in a dynamic transmission line and temporal photonic crystals, Appl Phys Lett **107**, (2015).

[21]  J. R. Zurita-Sánchez, J. H. Abundis-Patiño, and P. Halevi, Pulse propagation through a slab with time-periodic dielectric function ε(t), Opt. Express **20**, 5586 (2012).

[22]  J. R. Zurita-Sánchez, P. Halevi, and J. C. Cervantes-González, Reflection and transmission of a wave incident on a slab with a time-periodic dielectric function (t), Phys Rev A **79**, (2009).

[23]  A. M. Shaltout, J. Fang, A. V. Kildishev, and V. M. Shalaev, *Photonic Time-Crystals and Momentum Band-Gaps*, in *Conference on Lasers and Electro-Optics* (OSA, Washington, D.C., 2016), p. FM1D.4.

[24]  E. Galiffi, R. Tirole, S. Yin, H. Li, S. Vezzoli, P. A. Huidobro, M. G. Silveirinha, R. Sapienza, A. Alù, and J. B. Pendry, Photonics of Time-Varying Media, (2021).

[25]  A. Dikopoltseva, Y. Sharabia, M. Lyubarov, Y. Lumer, S. Tsesses, E. Lustig, I. Kaminer, and M. Segev, Light emission by free electrons in photonic time-crystals, (2022).

[26]  S. A. R. Horsley and J. B. Pendry, Quantum electrodynamics of time-varying gratings, Proc Natl Acad Sci U S A **120**, (2023).

[27]  M. Lyubarov, Y. Lumer, A. Dikopoltsev, E. Lustig, Y. Sharabi, and M. Segev, Amplified emission and lasing in photonic time crystals, Science (1979) **377**, 425 (2022).

[28]  F. R. Morgenthaler, Velocity Modulation of Electromagnetic Waves, IRE Transactions on Microwave Theory and Techniques **6**, 167 (1958).

[29]  S. E. Harris, J. E. Field, A. Imamoˇ, ; M Xiao, Y. Li, S. Jin, and J. Gea-Banacloche, Measurement of Dispersive Properties of Electromagnetically Induced Transparency in Rubidium Atoms, 1986.

[30]  J. T. Mendonça and P. K. K. Shukla, Physica Scripta Time Refraction and Time Reflection: Two Basic Concepts Time Refraction and Time Reflection: Two Basic Concepts, 2002.

[31]  F. Biancalana, A. Amann, A. V. Uskov, and E. P. O'Reilly, Dynamics of light propagation in spatiotemporal dielectric structures, Phys Rev E Stat Nonlin Soft Matter Phys **75**, (2007).

[32]  V. Bacot, M. Labousse, A. Eddi, M. Fink, and E. Fort, Time reversal and holography with spacetime transformations, Nat Phys **12**, 972 (2016).

[33]  H. Moussa, G. Xu, S. Yin, E. Galiffi, Y. Radi, and A. Alù, *Observation of Temporal Reflections and Broadband Frequency Translations at Photonic Time-Interfaces*.

[34]  T. R. Jones, A. V. Kildishev, M. Segev, and D. Peroulis, Time-Reflection of Microwaves by a Fast Optically-Controlled Time-Boundary, (2023).

[35]  O. Y. Long, K. Wang, A. Dutt, and S. Fan, Time reflection and refraction in synthetic frequency dimension, Phys Rev Res **5**, (2023).

[36] Z. Dong, H. Li, T. Wan, Q. Liang, Z. Yang, and B. Yan, Quantum time reflection and refraction of ultracold atoms, Nat Photonics (2023).

[37] D. Holberg and K. Kunz, Parametric properties of fields in a slab of time-varying permittivity, IEEE Trans Antennas Propag **14**, 183 (1966).

[38] A. B. Shvartsburg, Optics of nonstationary media, Physics-Uspekhi **48**, 797 (2005).

[39] F. Biancalana, A. Amann, and E. P. O'Reilly, Gap solitons in spatiotemporal photonic crystals, Phys Rev A (Coll Park) **77**, 011801 (2008).

[40] Y. Pan, M. I. Cohen, and M. Segev, Superluminal k -Gap Solitons in Nonlinear Photonic Time Crystals, Phys Rev Lett **130**, (2023).

[41] P. A. Franken, A. E. Hill, C. % Peters, and G. Weinreich, Generation of Optical Harmonics, 1961.

[42] J. A. Giordmaine, PHYSICAL REVIEW LETTERS MIXING OF LIGHT BEAMS IN CRYSTALS, 1962.

[43] P. D. Maker, R. % Terhune, M. Nisenoff, and C. M. Savage, EFFECTS OF DISPERSION AND FOCUSING ON THE PRODUCTION OF OPTICAL HARMONICS, 1962.

[44] J. A. Armstrong, N. Blgemeergen, J. Ducuing, and P. S. Pershan, Interactions between Light Waves in a Nonlinear Dielectric*, 1962.

[45] M. M. Fejer and G. A. Magel, Quasi-Phase-Matched Second Harmonic Generation: Tuning and Tolerances, 1992.

[46] A. Bahabad, M. M. Murnane, and H. C. Kapteyn, Quasi-phase-matching of momentum and energy in nonlinear optical processes, Nat Photonics **4**, 570 (2010).

[47] A. L. and G. E. Boyd Robert W. and Gaeta, *Nonlinear Optics*, in *Springer Handbook of Atomic, Molecular, and Optical Physics*, edited by G. W. F. Drake (Springer International Publishing, Cham, 2023), pp. 1097–1110.

[48] R. Tirole, S. Vezzoli, D. Saxena, S. Yang, T. V. Raziman, E. Galiffi, S. A. Maier, J. B. Pendry, and R. Sapienza, Second harmonic generation at a time-varying interface, Nature Communications **15**, (2024).

[49] S. Saha, S. Gurung, B. T. Diroll, S. Chakraborty, O. Segal, M. Segev, V. M. Shalaev, A. V Kildishev, A. Boltasseva, and R. D. Schaller, Third Harmonic Enhancement Harnessing Photoexcitation Unveils New Nonlinearities in Zinc Oxide. ArXivID: 2405.04891

[50] M. Segev, O. Segal, N. Konforty, M. Lyubarov, Y. Plotnik, E. Lustig, A. Dikopoltsev, Light-matter Interactions in Time-varying Media and in Photonic Time-crystals, International Conference on Ultrafast Phenomena, Barcelona, Spain, July 2024.